\documentclass[11pt]{article}

\usepackage{geometry, setspace}
\usepackage{hyperref}
\usepackage{natbib}

\usepackage{color, latexsym, amsfonts, amssymb, amsthm, amscd, amsmath, graphicx}
\usepackage{mathrsfs, bm}
\usepackage{float,multirow, booktabs} 

\newcommand{\be}{\begin{equation}}
\newcommand{\ee}{\end{equation}}
\newcommand{\bas}{\begin{eqnarray*}}
\newcommand{\eas}{\end{eqnarray*}}
\newcommand{\ba}{\begin{eqnarray}}
\newcommand{\ea}{\end{eqnarray}}
\newcommand{\bit}{\begin{itemize}}
\newcommand{\eit}{\end{itemize}}
\newcommand{\ben}{\begin{enumerate}}
\newcommand{\een}{\end{enumerate}}

\newcommand{\pr}{{\rm pr}}
\newcommand{\e}{ { \mathbb{E}}}
\newcommand{\diag}{\mbox{diag}}
\newcommand{\convergeto}{\overset{d}{\longrightarrow}}
\def\T{{ \mathrm{\scriptscriptstyle \top} }}

\newtheorem{theorem}{Theorem}
\newtheorem{remark}{Remark}

\begin{document}

\date{}
\title{
Maximum likelihood abundance estimation
 from 
 capture-recapture data
when covariates are missing at random
  }

\maketitle

\vspace{-0.6in}
\begin{center}
{Yang Liu and Yukun Liu \\
Key Laboratory of Advanced Theory and Application in Statistics and Data Science (East China Normal University), Ministry of Education \\
School of Statistics,
East China Normal University, Shanghai 200241, China  \\
liuyangecnu@163.com and  ykliu@sfs.ecnu.edu.cn
}

{Pengfei Li \\
Department of Statistics and Actuarial Science,
University of Waterloo, Waterloo, ON Canada N2L 3G1\\
{pengfei.li@uwaterloo.ca}
}

{Jing Qin\\
 National Institute of Allergy and Infectious Diseases,
 National Institute of Health, MD 20892, U.S.A.\\
 jingqin@niaid.nih.gov
}

{Lin Zhu\\
Institute of Statistics and Big Data, 
Renmin University of China, Beijing 100872, China\\
{linzhu\_2017@163.com}
}

\end{center}

\doublespacing

\begin{abstract}

In 
capture-recapture experiments,
individual covariates may be subject to missing,
especially when the number of times {of being captured} is small.
When the covariate information is missing at random,
the inverse probability weighting method and multiple imputation method
are widely used to obtain the point estimators of the abundance. 
These point estimators are then used to construct the 
Wald-type confidence intervals for the abundance.
However, such  intervals may have severely inaccurate
coverage probabilit{ies} and  their lower limits can be even less than
the number of individuals ever captured.
In this paper, 
we proposed a maximum empirical   likelihood estimation approach for the abundance
in presence of missing covariates. 
We show that  the maximum empirical likelihood estimator is asymptotically normal,
and that the empirical likelihood ratio statistic for abundance has a chisquare
limiting distribution with one degree of freedom.
Simulations indicate that
the proposed estimator has smaller mean square error
than the existing  estimators, 
and the proposed empirical likelihood ratio confidence interval
usually has more accurate coverage probabilities  than
the existing Wald-type confidence intervals.
We illustrate the proposed method by analyzing the bird species yellow-bellied prinia collected in Hong Kong.
\end{abstract}


\section{Introduction}

The estimation of population size or abundance is a fundamental problem
in many fields,  such as conservation biology,  demography, epidemiology
and software reliability 
\citep{tilling2001multilevel, borchers2002estimating, barnard2003using, boden2008capture}.
The population in abundance estimation  is usually assumed to be either closed or open.
A closed population  implies that  there is no birth,  death or migration
in the population and the abundance remains unchanged throughout the period of study.
Otherwise it is called an open population.
To estimate the abundance, 
the capture-recapture technique has been widely used to collect data,
given that a general census is usually too expensive or impractical.
In a capture-recapture experiment, individuals or animals from
the population were captured, marked or their existing marks noted if
they have been previously marked, and then released back into the population.
Depending on whether capture efforts are made  on separate  occasions or continuously,
the capture-recapture experiments can be divided into two types:
discrete time and continuous time.
In this paper, we focus on discrete time capture-recapture experiments, and 
make a closed population assumption so that the abundance
can be regarded as a parameter to be inferred.

As is widely accepted, heterogeneity is almost always present
in capture-recapture experiments  and ignoring
it may result in seriously biased estimates 
\citep{otis1978statistical, chao2001overview}.
To account for the heterogeneity,  \cite{huggins1989statistical}
 and \cite{alho1990logistic}
proposed to model the capture probability by a logistic regression model
on covariates such as individual characteristics and environmental conditions.
Since then, there have been a large body of literature on
abundance estimation  from capture-recapture data
when individual covariates are completely observed.
See 
\cite{chao2001overview, huggins2011review, stoklosa2011heterogeneous, liu2017maximum}
 and references therein.

In practice,  covariate information is vulnerable to missing.
For example, in real-life data from the Mai Po Sanctuary in Hong Kong
\citep{lin1999parametric, yip2002unified, wang2003semiparametric},
the determinant covariate `gender'  was missing for 74 of 132 captured
birds \citep{wang2005semiparametric}.
One more example can be found in Section 5. 
In this paper, we assume that the data are missing at random (MAR)
\citep{rubin1976inference, little2014statistical},
in other words, the missing of data  does not depend on the missing data themselves.
Under this assumption, 
\cite{xi2009estimation} assumed a parametric distribution
on the covariates and developed an EM-algorithm to obtain an
estimator of the population size.  To weaken the distribution assumption,
\cite{lee2016estimation} proposed three kinds of estimators through regression calibration,
inverse probability weighting and multiple imputation methods.
The latter two  are recommended because they are consistent.

The above reviewed estimation methods in the presence of missing covariate 
usually consists of three steps.
First,  the underlying parameters in the assumed models
are estimated by  solving certain estimating equations.
Second, a Horvitz-Thompson type estimator
is constructed for  the abundance  after the underlying parameters
are replaced by the obtained estimators in the first step.
Last,  Wald-type confidence intervals are then constructed
for interval estimation based on the asymptotic normality
of the Horvitz-Thompson type estimator.
These estimation methods often suffer from a number of
shortcomings.
The abundance estimator is often not efficient enough,
because the estimating equations are derived from conditional likelihood,
which is usually less efficient  than full likelihood.
Our simulation results also indicate that
the coverage accuracy of the Wald-type confidence interval
is usually unstatisfactory when the sample size is small or moderate.
In addition,  the lower limit of the Wald-type confidence interval
can be even less than the number of individuals captured.

When individual covariates are all completely observed,
\cite{liu2017maximum}  proposed a full empirical likelihood estimation
method for  the abundance.
The resulting maximum empirical likelihood estimator
and the empirical likelihood ratio confidence interval
were shown to outperform the traditional Horvitz-Thompson type  estimators
and Wald-type confidence intervals, which are based on conditional likelihood.
This method was further extended to continuous time capture-recapture data by
\cite{liu2018full}.
However if some individual covariates are missing,  
this method will produce  biased estimators
(see our simulation study)
if we simply discard  the subjects with missing covariates.

In this paper,  combining the conditional likelihood and
the celebrated empirical likelihood 
\citep{owen1988empirical, owen1990empirical},
we proposed a maximum full likelihood method for the abundance
in presence of missing covariates under the MAR assumption.
We show that   the maximum empirical likelihood estimator is asymptotically normal,
and that the empirical likelihood ratio statistic for abundance has a chisquare
limiting distribution with one degree of freedom.
The resulting empirical likelihood ratio confidence interval is one-step and free of variance estimation,
and its lower limit is always no less than {the number of times of being captured.}
Our  simulations indicate that
the proposed maximum likelihood estimator has smaller mean square error
than the existing Horvitz-Thompson type   estimators.
And the empirical likelihood ratio confidence interval
usually has more accurate coverage probabilities than
the Wald-type confidence intervals.

The rest of the paper is organized as follows.
In Section 2, we first introduce the capture probability model 
of \cite{huggins1989statistical} and \cite{alho1990logistic}
and consider the scenario that  the covariate information for some individuals are completely missing.
We then present our proposed method and investigate its large sample properties. 
As an extension, we further study the situation that some covariate information are completely observed for all individuals and 
the remaining are missing for some individuals. 
Section 3 compares the proposed method and  the existing methods through simulation studies. 
In Section 4, we illustrate the proposed method by analyzing the bird species
yellow-bellied prinia collected in Hong Kong.
Section 5 concludes the paper with a short discussion.

\section{Full Likelihood Estimation}

\subsection{Model and Data}

Let $\nu$ be the abundance of a closed population
and $K$  the number of capture occasions in a discrete time capture-recapture experiment.
For a generic individual in the population and $k=1, \ldots, K$,
define   $D_{k}^* = 1$ if it was captured on the $k$th occasion
and $D_{k}^* = 0$ otherwise.
Then   $(D_{1}^*, \dots, D_{K}^* )^\T$ is the so-called capture history.
Let   $Z^*$ be a $p$-dimensional covariate of this generic individual
with its first component being 1.
Following \cite{huggins1989statistical} and \cite{alho1990logistic},
we assume that  the capture probability on each occasion
follows the following logistic regression model or the Huggins-Alho model,
\ba
\label{logistic-model}
\pr(D_{k}^* = 1|Z^* = z) = \frac{\exp(\beta^\T z)}{1+\exp(\beta^\T z)} =: g(z;\beta).
\ea
Denote the number of times the generic individual being captured by  $D^* = \sum_{k = 1}^K D_{k}^*$.
Under the Huggins-Alho model, $D^*$ given $Z^*=z$ follows a binomial distribution with
size $K$ and  success probability $g(z;\beta)$.

In practice, individual covariates are vulnerable to missing if  $D^*$ is small.
Let $R^* $ be the missingness indicator with $R^*=1$ if the covariate of
a generic individual is not missing and $R^*=0$ otherwise.
In this paper, we adopt a MAR assumption, namely
the missing probability of a covariate is independent of the covariate itself,
i.e.,
\ba\label{MAR-model}
\pr(R^* = 1| Z^*=z, D^*=k ) = \pr(R^* = 1| D^*=k).
\ea

Suppose $n$ distinct individuals were captured at least once in the capture-recapture experiment.
Let $D_i $ and $Z_i$ respectively be the capture times and the covariate
of the $i$th individual ever captured.
Without of generality, we assume the first $m$ data are completely observed
and the covariates in the last $(n-m)$ data are missing.
Denoted by $R_i$ the corresponding missingness indicator, 
which equals 1 when $i = 1, \dots, m$ and 0 when $i = m+1,\dots,n$.
We wish to make inferences about the abundance $\nu$ below.

\subsection{Likelihood }

Suppose that conditional on $n$,
the observations
$
\{(R_i=1, D_i, Z_i): i = 1,\dots,m\}
\cup \{(R_i=0, D_i): i = m+1,\dots,n\}
$
are independent of each other.  Because all $D_i$'s satisfy $D_i>0$,
namely $\{D^*=D_i, D^*>0\} = \{D^* =D_i\}$,
it follows that for $1\leq i\leq m$
\bas
&& \pr(R^*=1, D^* = D_i, Z^* = Z_i | D^* >0) \\
&=&
\frac{ \pr(R^*=1, D^* = D_i, Z^* = Z_i)}{
\pr(  D^* >0)}  \\
&=&
\frac{ \pr(R^* = 1|D^* = D_i, Z^*=Z_i)\pr(D^* = D_i|Z^* = Z_i) \pr(Z^* = Z_i)}{
\pr(  D^* >0)} \\
&=&
\frac{ \pr(R^* = 1|D^* = D_i)\pr(D^* = D_i|Z^* = Z_i) \pr(Z^* = Z_i)}{
\pr(  D^* >0)},
\eas
where the last equality follows from the MAR assumption in \eqref{MAR-model}.
Similarly, for $m+1\leq i \leq n$
$$
\pr(R^*=0, D^* = D_i | D^* >0)
= \frac{ \pr(R^* =0| D^* = D_i) \pr(D^* = D_i)  }{\pr(  D^* >0)}.
$$

With the above preparations, we obtain  the full likelihood
\bas
&&  \pr( n )  \times \pr\{ (R_1=1, D_1, Z_1), \dots, (R_m=1, D_m, Z_m), \\
&& \hspace{4cm} (R_{m+1}=0, D_{m+1}), \dots, (R_n=0, D_n) | n \} \notag\\
   &=& \pr( n )  \times \prod_{i=1}^m \pr(R^*=1, D^* = D_i, Z^* = Z_i | D^* >0) \\
   && \hspace{4cm}
   \times \prod_{i=m+1}^n \pr(R^*=0, D^* = D_i | D^* >0) \\
  &=:& L_0\times L,
\eas
where
$
 L_0  =   \prod_{i = 1}^m   \pr(R^* = 1|D^* = D_i)
\times \prod_{i = m+1}^n \pr(R^* =0| D^* = D_i)
$
and
\bas
L
  &=&  \pr( n ) \times \prod_{i = 1}^m \{ \pr(D^* = D_i|Z^* = Z_i) \pr(Z^* = Z_i) \} \\
&& \hspace{3cm}
\times \prod_{i = m+1}^n  \pr(D^* = D_i)  \times \{\pr(D^* > 0) \}^{-n}.
\eas
We note that   $ L_0$  is the likelihood contribution of $\{(R_i, D_i): i=1, \ldots, n\}$.
Since it does not involve the main parameter of interest $\nu$,
we abandon it and proceed with only the partial likelihood $L$.

For $k=0, \ldots, K$, let  $\gamma_k = \pr(D^* = k)$ and
$\alpha = (\gamma_0, \gamma_1, \dots, \gamma_K)^\T$.
Clearly, $\pr(D^*>0) = 1-\gamma_0$ and
$n$ follows the binomial distribution B$(\nu, 1-\gamma_0)$, i.e.,
\ba\label{like1}
\pr( n ) = { \nu \choose n }(1 - \gamma_0)^n \gamma_0^{\nu - n}.
\ea
For $k=1,\dots,K$,  let $m_k$ be the number of individuals
which were captured exactly $k$ times and whose covarites were missing.
Then
\ba\label{like2}
\prod_{i=m+1}^n \pr(D^* = D_i) =   \prod_{k=1}^K \gamma_k^{m_k}.
\ea
It follows from  the Huggins-Alho model  that
\ba\label{like3}
\prod_{i = 1}^m    \pr(D^* = D_i|Z^* = Z_i)   =
\prod_{i = 1}^m {K \choose D_i}\{g(Z_i; \beta)\}^{D_i} \{1 - g(Z_i; \beta)\}^{K - D_i}.
\ea
Putting Equations (\ref{like1}) - (\ref{like3})
into   $L$
gives
$$
L =
{ \nu \choose n } \gamma_0^{\nu-n} \times
\prod_{i = 1}^m \{g(Z_i; \beta)\}^{D_i} \{1 - g(Z_i; \beta)\}^{K - D_i} \times
  \prod_{k=1}^K \gamma_k^{m_k} \times
\prod_{i = 1}^m \pr(Z^* = Z_i).
$$

{Let $F_{Z^*}$ and ${\rm d} F_{Z^*}$ be the cumulative distribution 
and the probability density or mass function of $Z^*$, respectively. 
That is, $\pr(Z^* = z) = {\rm d} F_{Z^*} (z)$.
Below, we propose to handle the marginal distribution $F_{Z^*}$ by empirical likelihood 
\citep{owen1988empirical, owen1990empirical}.
}
Let $U(z; \alpha, \beta) = (u_{0}(z; \gamma_0, \beta), \dots, u_K(z; \gamma_K, \beta) )^\T$ with
$$
 u_k(z; \gamma_k, \beta) = {K \choose k} \{g(z; \beta)\}^{k} \{1 - g(z; \beta)\}^{K - k} - \gamma_k, \; k = 0, 1,\dots, K.
$$
It can be verified that
\ba\label{eq-constraints}
\int U(z; \alpha, \beta) {\rm d}F_{Z^*}(z) = 0,
\ea
when $(\alpha, \beta)$ takes its true value.
By the principle of empirical likelihood,
we discrete the distribution of $Z^*$ by
\bas
F_{Z^*}(z) = \sum_{i = 1}^m p_i I(Z_i \leq z).
\eas
Then the empirical log-likelihood  is
$$
\log{\nu \choose n} + \sum_{k = 0}^K m_k \log (\gamma_k) + \sum_{i = 1}^m\log (p_i)
 + \sum_{i = 1}^m \Big[
D_{i} \log\{g(Z_i; \beta)\} + (K- D_{i})\log\{1 - g(Z_i; \beta)\} \Big],
$$
where $m_0=\nu-n$ and
the feasible $p_i$'s satisfy
$$
p_i \geq 0, \; \sum_{i = 1}^m p_i = 1, \; \sum_{i = 1}^m U(Z_i; \alpha, \beta)p_i = 0.
$$

For a given $(\alpha, \beta)$, the empirical log-likelihood achieves its maximum in general when
$$
p_i = \frac{1}{ m} \frac{1}{ \{ 1 + \lambda^\T U(Z_i; \alpha, \beta) \} },
$$
where the Lagrange multiplier $\lambda = (\lambda_0, \dots, \lambda_K)^\T$ is the solution of
\ba\label{equ-lambda}
\sum_{i = 1}^m \frac{U(Z_i; \alpha, \beta)}{1 + \lambda^\T U(Z_i; \alpha, \beta)} = 0.
\ea
Profiling out $p_i$'s, we obtain the profile empirical log-likelihood of $(\nu, \alpha, \beta)$
\bas
\ell(\nu, \alpha, \beta) &=&
\log{\nu \choose n} +
\sum_{k = 0}^K m_k \log (\gamma_k) - \sum_{i = 1}^m\log\{ 1 + \lambda^\T U(Z_i; \alpha, \beta ) \} \notag \\
&+ &\sum_{i = 1}^m \Big[
D_{i} \log\{g(Z_i; \beta)\} + (K- D_{i})\log\{1 - g(Z_i; \beta)\} \Big],
\eas
where $\lambda = \lambda(\alpha,\beta)$ satisfies Equation (\ref{equ-lambda}).

\subsection{Estimation and asymptotics}
Given the   profile empirical log-likelihood
$
\ell(\nu, \alpha, \beta)
$,
we propose to estimate the parameters by
their  maximum empirical likelihood estimator
$$
(\hat\nu, \hat\alpha, \hat\beta) = \arg\max_{(\nu, \alpha, \beta)} \ell (\nu, \alpha, \beta).
$$
Accordingly we define the empirical likelihood ratio functions of $(\nu, \alpha, \beta)$ and $\nu$ as
\bas
R(\nu, \alpha, \beta)  &=&
2\{\max_{(\nu, \alpha, \beta)}\ell(\nu, \alpha, \beta) - \ell(\nu, \alpha, \beta)\}
= 2\{\ell(\hat\nu, \hat\alpha, \hat\beta) - \ell(\nu, \alpha, \beta)\}, \\
R'(\nu)  &=&
2\{\max_{(\nu, \alpha, \beta)}\ell(\nu, \alpha, \beta) - \max_{(\alpha, \beta)}\ell(\nu, \alpha, \beta)\}
= 2\{\ell(\hat\nu, \hat\alpha, \hat\beta) - \ell(\nu, \hat\alpha_\nu, \hat\beta_\nu)\},
\eas
where $ (\hat\alpha_\nu, \hat\beta_\nu) = \arg\max_{(\alpha, \beta)} \ell(\nu, \alpha, \beta)$ given $\nu$.

The theorem below shows that the maximum empirical likelihood estimator
is asymptotically normal
and the two empirical likelihood ratio statistics
follow asymptotically chi-square distributions.
We define necessary notation to ease the exposition.
Let
  $(\nu_0, \alpha_0, \beta_0)$ be the true value of $(\nu, \alpha, \beta)$ and
$\alpha_0 = (\gamma_{00}, \dots, \gamma_{K0})$ with elements in $(0,1)$.
Define $h_k =\pr(R^* = 1| D^*=k)$,
$\lambda_{00} = - (  \sum_{k = 1}^K h_k \gamma_{k0} )^{-1}, {H}_1= (1, 1-h_1, \dots, 1-h_K)^\T$, and
${H}_2 = \diag\{ 1/\gamma_{00} , (1-h_1)/\gamma_{10}, \dots,  (1-h_K)/\gamma_{K0} \}$. 
Let  $\pi(z;\beta)$ be the probability that an individual has been captured at least once 
with no missing covariate for certain $Z^*=z$.
That is,
$\pi(z;\beta) = \sum_{k=1}^K {K\choose k} \{g(z;\beta)\}^k\{1 - g(z;\beta)\}^{K-k} h_k$.
Denote  the first and second derivatives of $\pi(z;\beta)$ with respect to $\beta$
  by $\dot\pi(z;\beta)$ and $\ddot\pi(z;\beta)$, respectively.
Let $\e$ be the expectation operator  with respect to $F_{Z^*}$, and
  $A^{\otimes 2} = AA^\T$ for  a vector or matrix $A$.
In addition, we use ${0}_{K\times1}$ and $I_{K+1}$
to denote a $K$-order vector with elements zero
and  the $(K+1)$-order identity matrix.
The following matrix is very important in our theoretical study:
\ba\label{equ-W}
   W&=&\left(
     \begin{array}{ccc}
       V_{11} & V_{12}&0_{K\times1}^\T\\
       V_{21} & V_{22}- V_{24}V_{44}^{-1} V_{42} & V_{23}- V_{24}V_{44}^{-1} V_{43} \\
       0_{K\times1} & V_{32}- V_{34}V_{44}^{-1} V_{42} & V_{33}- V_{34}V_{44}^{-1} V_{43}
     \end{array}
   \right),
\ea
where $V_{11} = \gamma_{00}^{-1} - 1, 
V_{21} = V_{12}^\T = (-\gamma_{00}^{-1}, {0}_{K\times1}^\T)^\T$, 
and
\bas
V_{22} &=& H_2  - \e \left\{ \frac{1}{\pi(Z^*; \beta_0)} \right\} \times H_1^{\otimes2}, \quad
V_{32} = V_{23}^\T = - \e \left\{ \frac{\dot\pi(Z^*;\beta_0)}{\pi(Z^*; \beta_0)}\right\} \times H_1^\T, \\
V_{42} &=& V_{24}^\T =
	\lambda_{00}^{-1} I_{K+1} +
	\lambda_{00}^{-1} \e\left\{
	\frac{ U(Z^*; \alpha_0, \beta_0 ) }{ \pi(Z^*; \beta_0) } \right\} \times  H_1^\T, \\
V_{33} &=& \e\left[  \ddot\pi(Z^*;\beta_0)  -
	\frac{ \{\dot\pi(Z^*;\beta_0)\}^{\otimes 2}  }{ \pi(Z^*;\beta_0) }
	 + K g(Z^*; \beta_0)\{ 1 - g(Z^*; \beta_0) \} \pi(Z^*;\beta_0)Z^{*\otimes 2}
	\right], \\
V_{34}&=& V_{43}^\T = - \lambda_{00}^{-1} \e\left\{ \frac{\partial U^\T(Z^*; \alpha_0, \beta_0 ) }{\partial\beta} -
	\frac{ \dot\pi(Z^*; \beta_0) U^\T(Z^*; \alpha_0, \beta_0 )}{\pi(Z^*; \beta_0)}
	\right\}, \\
V_{44}&=& - \lambda_{00}^{-2}
	\e \left\{ \frac{ U(Z^*;\alpha_0, \beta_0) U^\T(Z^*;\alpha_0, \beta_0) }{
	\pi(Z^*;\beta_0)} \right\}.
\eas

\begin{theorem}
\label{asy-pro}
Suppose $\int \{ \pi(z; \beta)\}^{-1}{\rm d}F_{Z^*}(z)<\infty$ for $\beta$ in a neighborhood of $\beta_0$.
If the matrix $W$ defined in Equation \eqref{equ-W} is nonsingular,   then as $\nu_0$ goes to infinity,
\bit
\item[(a)]
$\sqrt{\nu_0} \left\{\log(\hat \nu/\nu_0),  \;
\hat \alpha - \alpha_0, \;
\hat \beta - \beta_0 \right\}^{\T}
\convergeto   N(0, W^{-1})$,  where $\convergeto$ stands for convergence in distribution;
\item[(b)]
$R(\nu_0, \alpha_0, \beta_0)\convergeto \chi_{K+p+2}^2$
and
$R'(\nu_0)  \convergeto \chi_1^2$.
\eit
\end{theorem}

According to Theorem \ref{asy-pro}, the proposed likelihood ratio
confidence interval for $\nu$
\ba
\label{lrt-ci}
\mathcal{I} = \{\nu:  R'(\nu) \leq \chi^2_{1}(1-a)\}
\ea
where $\chi^2_{1}(1-a)$ is the $(1-a)$-quantile of $\chi_1^2$
has an asymptotically correct coverage probability when the confidence level is $1-a$.
Compared with the usual Wald-type confidence interval,
the proposed interval is clearly free from variance estimation.
Meanwhile its lower limit is never less than $n$, as the domain of the likelihood ratio function
$R'(\nu)$ is $[n, \infty)$.

\begin{remark}
We have assumed that $\gamma_{k0} \in (0, 1)$ for $k=0,1,\ldots, K$
or $\min_{0\leq k\leq K}\gamma_{k0}>0$.
It then follows that  $\min_{0\leq k\leq K} m_k >0$ holds with probability
approaching 1 as $\nu_0\rightarrow \infty$.
This implies that   the number of constraints
in Equation (\ref{eq-constraints}) is exactly equal to $K+1$
as $\nu_0$ is large.
However this is not always the case in practice,
especially when $n$ is small.
If some $m_k$ are zero, we  simply ignore the corresponding estimating
equations, and proceed the maximization as usual.
In this situation,  although the maximum likelihood
estimators of the corresponding  $\gamma_{k}$'s have no definition,
the asymptotic results in Theorem   \ref{asy-pro} are still valid.
The validation of the proposed confidence interval in \eqref{lrt-ci}
is therefore  still  guaranteed.
\end{remark}

\subsection{An extension}
In the previous subsection, we assume that the probability of a generic individual
being captured in one capture occasion is affected only by $D^*$,
the number of times that an individual was captured.
Motivated by the yellow-bellied prinia data analysis in Section 5,
we extend Model \eqref {MAR-model}
to incorporate a binary covariate (such as fat index in the real data)
which is completely observed.
Let $Z^*=(1, X^*, Y^*)$, where $X^*$ is the always-observed binary covariate  and $Y^*$
the covariate subject to missing. We assume
$$
\pr(R^* = 1| Z^*=z,   D^*=k ) = \pr(R^* = 1| X^*=x, D^*=k)
$$
for all $x=0, 1$ and $ k=0, 1, \ldots, K. $

In this situation, the observations are
$
\{(D_i, X_i, Y_i): i = 1,\dots,m\}
$
and
$\{(D_i, X_i): i = m+1,\dots,n\},
$
where $X_i, Y_i, D_i$  are the realizations of  $X^*, Y^*, D^*$ if they can be observed.
By similar derivations to the previous subsection,
we have the partial likelihood in the current situation
\bas
L_e
&=&  \pr( n ) \times \prod_{i=1}^m   \{\pr( D^*=D_i|Z^* = Z_i) \pr( Z^* = Z_i)\} \\
&&
\times \prod_{i= m+1}^n  \pr( X^* = X_i, D^*=D_i) \times \{\pr(D^*>0)\}^{-n}.
\eas

Let $\alpha = (\gamma_0, \gamma_{01}, \dots, \gamma_{0K}, \gamma_{11},\dots,\gamma_{1K})^\T$
with $\gamma_0 = \pr(D^* = 0)$ and $\gamma_{jk} = \pr(X^*=j, D^* = k)$ for $j=0,1$ and $k=1,\ldots,K$.
Let $m_{jk}$ be the number of individuals being captured exactly $k$ times with $Y^*$ missing and $X^*=j$.
Clearly
$$
\prod_{i=m+1}^n \pr(X^*=X_i, D^* = D_i) = \prod_{j=0}^1 \prod_{k=1}^K \gamma_{jk}^{m_{jk}}.
$$
Accordingly we have the partial likelihood
$$
L_e  =
{ \nu \choose n } \gamma_0^{\nu-n} \times
\prod_{i = 1}^m \{g(Z_i; \beta)\}^{D_i} \{1 - g(Z_i; \beta)\}^{K - D_i} \times
\prod_{j=0}^1 \prod_{k=1}^K \gamma_{jk}^{m_{jk}} \times
\prod_{i = 1}^m \pr(Z^* = Z_i),
$$
where the function $g(\cdot)$ is given in \eqref{logistic-model}.

Let   $z=(1, x, y)^\T$ and $I(\cdot)$ be the indicator function.
Define $U(z; \alpha, \beta)$ to
be the vector  consisting of the following
estimating equations
\bas
U_{0}(z;\alpha,\beta) &=&
\{1 - g(z;\beta)\}^K - \gamma_{0}, \\
U_{jk}(z;\alpha,\beta) &=& {K\choose k}  \{g(z;\beta)\}^{k} \{1 - g(z;\beta)\}^{K-k}I(x=j) -\gamma_{jk},
\eas
for $ j = 0,1$ and $ k = 1,\ldots, K$.
Again it can be verified that
$\int U(z; \alpha, \beta) dF_{Z^*}(z) = 0$
when $(\alpha, \beta)$ takes its true value.

With the above preparations,  up to a constant, the empirical log-likelihood
is
\bas
\log{\nu \choose n} + (\nu-n)\log(\gamma_0) +
\sum_{j = 0}^1 \sum_{k = 1}^K m_{jk} \log (\gamma_{jk}) + \sum_{i = 1}^m\log (p_i) \\
 + \sum_{i = 1}^m \Big[
D_{i} \log\{g(Z_i; \beta)\} + (K- D_{i})\log\{1 - g(Z_i; \beta)\} \Big],
\eas
where the feasible $p_i$'s satisfy
$$
p_i \geq 0, \; \sum_{i = 1}^m p_i = 1, \; \sum_{i = 1}^m U(Z_i; \alpha, \beta)p_i = 0.
$$
Profiling $p_i$'s out,
we obtain the profile empirical log-likelihood
\bas
\ell_e (\nu, \alpha, \beta) &=&
\log {\nu \choose n} + (\nu - n)\log(\gamma_{0})
+ \sum_{j = 0}^1\sum_{k=1}^K m_{jk}\log(\gamma_{jk})
 - \sum_{i = 1}^m\log\{ 1 + \lambda^\T U(Z_i; \alpha, \beta) \} \\
&&  + \sum_{i=1}^{m}[ D_i\log\{g(Z_i; \beta)\} + (K-D_i)\log\{1-g(Z_i; \beta)\}],
\eas
where
$\lambda = \lambda(\alpha, \beta)$ is the solution of
$$
\sum_{i = 1}^m \frac{U(Z_i; \alpha, \beta)}{1 + \lambda^\T U(Z_i; \alpha, \beta)} = 0.
$$

Denote the maximum empirical likelihood estimator by
$
(\hat\nu_e, \hat\alpha_e, \hat\beta_e) = \arg\max_{(\nu, \alpha, \beta)} \ell_e (\nu, \alpha, \beta),
$
and define the empirical likelihood ratio functions of $(\nu, \alpha, \beta)$ and $\nu$ as
$$
R_e(\nu, \alpha, \beta)  =
2\{\max_{(\nu, \alpha, \beta)}\ell_e(\nu, \alpha, \beta) - \ell_e(\nu, \alpha, \beta)\}, \quad
R_e'(\nu)  =
2\{\max_{(\nu, \alpha, \beta)}\ell_e(\nu, \alpha, \beta) - \max_{(\alpha, \beta)}\ell_e(\nu, \alpha, \beta)\}.
$$
Let $\alpha_0 = (\gamma_{00}, \gamma_{010},\dots,\gamma_{0K0},\gamma_{110},\dots, \gamma_{1K0})$
be the true value of $\alpha$.
Define $h_{jk}=\pr(R^* = 1| X^*=j, D^*=k)$,
$\lambda_{00 }
= -\{ \sum_{j=0}^1 \sum_{k=1}^K  h_{jk} \gamma_{jk0} \}^{-1}$, 
${H}_{1e}= (1, 1-h_{01}, \dots,1-h_{0K}, 1-h_{11}, \dots, 1-h_{1K})^\T$, 
$
{H}_{2e} = \diag\{1/\gamma_{00}, (1-h_{01})/\gamma_{100}, \dots, (1-h_{0K})/\gamma_{0K0}, 
(1-h_{11})/\gamma_{110}, \dots, (1-h_{1K})/\gamma_{1K0}\}
$,
and
$$
\pi_e(z; \beta) = \pr(R^* = 1, D^*>0 | Z^* = z)
= \sum_{j=0}^1 \sum_{k=1}^K {K\choose k} \{g(z;\beta)\}^k\{1 - g(z;\beta)\}^{K-k} I(x=j) h_{jk}.
$$
Besides, we define $W_e$ to be the $W$ in \eqref{equ-W}  except that
$V_{21} = V_{12}^\T$ becomes $(-\gamma_{00}^{-1}, {0}_{(2K)\times1}^\T)^\T$, and
$0_{K+1}, H_1, H_2$ and $\pi(\cdot)$ are respectively replaced by
$0_{2K+1}, H_{1e}, H_{2e}$ and $\pi_e(\cdot)$.
With these notation introduced above we have the following theorem.

\begin{theorem}
\label{asy-pro-2}
Suppose $\int \{ \pi_e(z; \beta)\}^{-1} {\rm d}F_{Z^*}(z)<\infty$ for $\beta$ in a neighborhood of $\beta_0$.
If the matrix $W_e$ is nonsingular,   then as $\nu_0$ goes to infinity,
\begin{enumerate}
\item[(a)]
$\sqrt{\nu_0} \left\{\log(\hat \nu_e/\nu_0),  \;
\hat \alpha_e - \alpha_0, \;
\hat \beta_e - \beta_0 \right\}^{\T}
\convergeto   N(0, W_e^{-1})$;
\item[(b)]
$R_e(\nu_0, \alpha_0, \beta_0)\convergeto \chi_{2K+p+2}^2$
and
$R'_e(\nu_0)  \convergeto \chi_1^2$.
\end{enumerate}
\end{theorem}

Theorem \ref{asy-pro-2} indicates again that the empirical likelihood ratio confidence interval
$$
\mathcal{I}_e = \{ \nu: R'_e(\nu)\leq \chi^2_{1,1-a}\}
$$
has an asymptotically correct coverage probability when the confidence level is $(1-a)$.
It is worth noting that we take only one category variable into account
in the extension for ease of presentation.
The proposed maximum likelihood estimation method
can naturally be extended to incorporate more categorical variables.

\section{Simulation}
In this section,
we investigate   the finite sample performance of the proposed estimation methods
(abundance estimator $\hat \nu$ or $\hat \nu_e$ and confidence interval $\mathcal{I}$ or $\mathcal{I}_e$)
by comparing them  with  the methods of 
\cite{lee2016estimation} and \cite{liu2017maximum}.
Let $\tilde\nu_1$ and $\tilde\nu_2$ be the inverse probability weighting estimator $\hat N_w$ and
the multiple imputation estimator $\hat N_{m2}$
of  \cite{lee2016estimation}
with variance estimates $\tilde\sigma_1^2$ and $\tilde\sigma_2^2$,  respectively.
Let $\mathcal{I}_1$ and $\mathcal{I}_2$  be their corresponding Wald-type confidence intervals,
that is,
$$
\mathcal{I}_1 = \{\nu: (\tilde\nu_1 - \nu)^2/\tilde\sigma_1^2 \leq \chi^2_{1,1-a} \}\quad
\text{and }\quad
\mathcal{I}_2 = \{\nu: (\tilde\nu_2 - \nu)^2/\tilde\sigma_2^2 \leq \chi^2_{1,1-a} \}.
$$
We may simply abandon the data with missing values and apply
the maximum likelihood estimation method of \cite{liu2017maximum}, which
corresponds to the complete case in the usual missing data problem.
Let $\tilde\nu_3$ be the maximum empirical likelihood abundance estimator $\hat N_s$
and $\mathcal{I}_3$  the corresponding likelihood ratio confidence interval
$\mathcal{I}_{1s}$ in \cite{liu2017maximum}.

For a generic abundance estimator
$\breve\nu$,
its bias and relative mean squared error (RMSE) are defined as
$$
\rm Bias(\breve\nu) =  \e \breve\nu - \nu_0
\quad\text{and}\quad
\rm RMSE(\breve\nu) = \e (\breve\nu - \nu_0)^2/\nu_0
$$
and taken as
criteria to   evaluate its finite-sample performance.
For a generic two-sided confidence interval $[\nu_L, \nu_U]$,
accompanying are two one-sided intervals
and $[\nu_L, \infty]$ (Lower limit) and $[n, \nu_U]$ (Upper limit).
We compare all interval estimators by their coverage probability.

%

\subsection{Simulation set-up}
We generate data from the following four scenarios:
\bit
\item[A.]
Set the total number of capture occasions $K = 2$
and  individual covariate $Z^* = (1, Y^*)$,
where $Y^*$ follows the uniform distribution  U(0,\ 3).
Given $Z^*=z$, we generate  $D^*$ from a Binomial distribution  B$(K,\; g(z; \beta))$,
where  the true value of $\beta$ is $(-2, 1)^\T$.
The selection probability defined in Equation (\ref{MAR-model}) is set to
$\pr(R^*=1|D^*=k) = \{1+\exp(0.5 -0.7k)\}^{-1}$.
Under this setting, the overall capture probability and
 the missing rate are about 60\% and 40\%, respectively.

\item[B.]
The same as Scenario A except  $K=5$.
The overall capture probability and the missing rate become around 84\% and 27\%, respectively.

\item[C.]
We consider the individual covariate $Z^*=(1, X^*, Y^*)$
with $X^*\sim {\rm B}(1,\;  0.3)$,
and $Y^*\sim {\rm U}(0,\; 3)$.
Given $Z^*=z$, we generate $D^*$
from B$(K,\; g(z;\beta))$
where $K=2$
and the true value of $\beta$ is $(-2, 1, 1)^\T$.

Here  $X^*$'s are completely observed and
$Y^*$'s are subject to missing with
the selection probability
$h(x,k) = \{1+\exp( 0.5  - 0.7x - 0.7k)\}^{-1}$.
Under this setting,
the overall capture probability and the missing rate are about 66\% and 34\%, respectively.

\item[D.]
The same as Scenario C except  $K=5$.
The overall capture probability and the missing rate become around 88\% or 22\%, respectively.

\eit

In each scenario, we set the population size to be $\nu_0 = 200$ and $400$.
All the simulation results reported are obtained based on 5000 simulation repetitions.

\subsection{Simulation results}

{\it Comparison of point estimation} \quad
Table \ref{table-point-est} tabulates
the biases and the relative mean square errors of
the four abundance estimators in comparison.
We see that the proposed estimator $\hat \nu$ or $\hat \nu_e$
has uniformly smaller bias and RMSE than
\cite{lee2016estimation}'s two estimators $\tilde \nu_1$ and $\tilde \nu_3$,
although all of them have acceptable performance.

To our surprise, ignoring all missing data,
 the estimator $\tilde\nu_3$ is severely biased,
in particular when the true abundance $\nu_0$ increases.
This is very different from the usual missing data problem,
in which  the complete-case estimator is still consistent although
it may not be efficient under the MAR assumption.
This phenomenon was also observed by \cite{lee2016estimation}.

\begin{table*}[h]
\centering
\caption{\label{table-point-est}
Biases and RMSEs of the proposed estimator  $\hat \nu$ (or $\hat\nu_e$) and
three existing estimators $\tilde\nu_1$,
$\tilde\nu_2$, and $\tilde \nu_3$.
 }

\begin{tabular}{cccccccccccccccccc}
\toprule
& & \multicolumn{4}{c}{Scenario A} & \multicolumn{4}{c}{Scenario B} \\

\cmidrule(r){3-6} \cmidrule(r){7-10}

&$\nu_0$	&$\hat\nu$ 	&$\tilde\nu_1$ 	&$\tilde\nu_2$ 	&$\tilde\nu_3$ 
		&$\hat\nu$ 	&$\tilde\nu_1$ 	&$\tilde\nu_2$ 	&$\tilde\nu_3$ \\
\midrule

\multirow{2}{*}{Bias}	&200	 	&25.37	&34.58	&34.81	&-83.81  
						&0.08 	&1.84	&1.85	&-62.99 \\
				&400		&16.15	&23.91	&24.10	&-181.48           
						&-0.21	&1.49	&1.51	&-126.04 \\
\multirow{2}{*}{RMSE}&200	&64.94	&72.68	&72.12	&47.09          
						&0.63	&0.70	&0.70	&20.25 \\
				&400		&16.56	&19.33	&19.75	&85.76    
						&0.61	&0.64	&0.65	&40.11 \\
\midrule
& & \multicolumn{4}{c}{Scenario C} & \multicolumn{4}{c}{Scenario D}  \\

\cmidrule(r){3-6} \cmidrule(r){7-10}

&$\nu_0$	&$\hat\nu_e$ 	&$\tilde\nu_1$ 	&$\tilde\nu_2$ 	&$\tilde\nu_3$ 
		&$\hat\nu_e$ 	&$\tilde\nu_1$ 	&$\tilde\nu_2$ 	&$\tilde\nu_3$ \\
\midrule
\multirow{2}{*}{Bias}	&200	 	&14.15	&21.44	&22.04	&-73.14 
						&-0.06	&1.44	&1.46	&-51.19 \\
				&400		&10.55	&16.44	&16.72	&-158.00
						&-0.26 	&1.20	&1.22	&-102.26 \\
\multirow{2}{*}{RMSE}&200	&19.27	&23.93	&26.25	&35.89 
						&0.42	&0.46	&0.46	&13.44\\
				&400		&8.99	&10.17	&10.30	&64.64   
						&0.38	&0.40	&0.40	&26.47\\
\bottomrule
\end{tabular}
\end{table*}

\begin{table*}[h]
\scriptsize
\centering
\caption{\label{table-conf}	Simulated coverage probabilities of
the proposed confidence interval  ($\mathcal{I}$ or $\mathcal{I}_e$),  and
  the two Wald-type confidence intervals ($\mathcal{I}_1$ and $\mathcal{I}_2$).
 }
\begin{tabular}{cccccccccccccccccc}
\toprule
&&	& \multicolumn{3}{c}{Scenario A} & \multicolumn{3}{c}{Scenario B} 
	& \multicolumn{3}{c}{Scenario C} & \multicolumn{3}{c}{Scenario D}  \\

\cmidrule(r){4-6} \cmidrule(r){7-9} \cmidrule(r){10-12} \cmidrule(r){13-15}
Type&Level&$\nu_0$	
		&$\mathcal{I}$ 	&$\mathcal{I}_1$ 	&$\mathcal{I}_2$
		&$\mathcal{I}$ 	&$\mathcal{I}_1$	&$\mathcal{I}_2$
		&$\mathcal{I}_e$&$\mathcal{I}_1$ 	&$\mathcal{I}_2$
		&$\mathcal{I}_e$&$\mathcal{I}_1$ 	&$\mathcal{I}_2$ \\
\midrule
\multirow{6}{*}{Two-sided}		&\multirow{2}{*}{90\%}	&200	& 
89.24	& 89.94	&89.24	&89.72	&91.20	&90.84	&89.76	&91.96	&91.20	&89.16	&90.40	&90.06\\
						&					&400	&
89.94	& 91.04	&90.04	&90.02	&90.98	&90.34	&89.88	&91.82	&91.04	&90.52	&91.08	&90.88\\
						&\multirow{2}{*}{95\%}	&200	&
94.68	&92.50	&92.02	&94.74	&95.16	&94.66	&94.58	&94.18	&93.46	&94.76	&94.54	&94.52\\
						&					&400	& 
94.86	&93.60	&93.02	&95.10	&95.32	&94.92	&94.96	&94.58	&93.98	&95.36	&95.46	&95.36\\
						&\multirow{2}{*}{99\%}	&200	& 
98.68	&95.92	&95.22	&99.06	&97.78	&97.72	&98.94	&96.80	&96.28	&98.92	&97.86	&97.82\\
						&					&400	& 
98.88	&96.70	&96.24	&99.02	&98.42	&98.24	&98.90	&97.54	&97.00	&99.14	&98.68	&98.58\\
\midrule
\multirow{6}{*}{Lower limit}	&\multirow{2}{*}{90\%}	&200	& 
89.10	&85.70	&84.90	&88.78	&88.26	&87.80	&90.48	&87.78	&86.94	&88.20	&88.00	&87.68\\
						&					&400	& 
89.46	&87.02	&85.62	&88.62	&88.22	&87.92	&89.60	&87.50	&86.76	&89.06	&89.00	&88.72\\
						&\multirow{2}{*}{95\%}	&200	& 
94.46	&89.94	&89.24	&94.22	&92.92	&92.68	&95.16	&91.96	&91.20	&93.46	&92.44	&92.18\\
						&					&400	& 
94.70	&91.04	&90.04	&94.46	&93.48	&93.02	&94.88	&91.88	&91.18	&94.60	&93.74	&93.92\\
						&\multirow{2}{*}{99\%}	&200	& 
98.84	&94.66	&94.18	&98.74	&96.96	&96.88	&98.92	&95.92	&95.22	&98.68	&97.02	&96.86\\
						&					&400	& 
98.94	&95.82	&95.16	&98.78	&97.72	&97.54	&99.02	&96.60	&96.04	&98.84	&97.92	&97.96\\
\midrule
\multirow{6}{*}{Upper limit}	&\multirow{2}{*}{90\%}	&200	& 
89.76	&100.0	&99.86	&91.36	&93.28	&93.14	&89.56	&99.74	&99.38	&91.08	&92.98	&92.78\\
						&					&400	& 
90.92	&99.12	&98.06	&90.88	&92.30	&92.22	&89.88	&97.56	&96.02	&91.46	&92.78	&92.44\\
						&\multirow{2}{*}{95\%}	&200	& 
94.78	&100.0	&100.0	&95.50	&98.28	&98.16	&94.60	&100.0	&100.0	&95.70	&97.96	&97.88\\
						&					&400	& 
95.24	&100.0	&100.0	&95.56	&97.50	&97.32	&95.00	&99.94	&99.86	&95.92	&97.34	&97.28\\
						&\multirow{2}{*}{99\%}	&200	& 
98.82	&100.0	&100.0	&99.26	&99.96	&99.98	&98.78	&100.0	&100.0 	&99.08	&99.98	&100.0\\
						&					&400	& 
98.88	&100.0	&100.0	&99.16	&99.96	&99.96	&99.16	&100.0	&100.0 	&99.16	&99.94	&99.94\\
\bottomrule
\end{tabular}
\end{table*}

{\it Comparison of interval estimation} \quad
Table \ref{table-conf} reports the two-sided and one-sided
coverage probabilities of  the proposed interval ($\mathcal{I}$ or $\mathcal{I}_e$)
and the two Wald-type intervals  ($\mathcal{I}_1$ and $\mathcal{I}_2$)
when the nominal levels are  90\%, 95\% and 99\%, respectively.  Because
the corresponding estimator $\tilde \nu_3$ is severely biased,
the complete-case interval estimator $\mathcal{I}_3$ is expected
to have poor coverage probabilities and is hence omitted.

Clearly, we can see that  the proposed confidence interval $\mathcal{I}$ or $\mathcal{I}_e$
always has the best performance among the three intervals
in terms of both one-sided and two-sided coverage accuracies.
For two-sided confidence intervals, 
although the two Wald-type confidence intervals $\mathcal{I}_2$ and $\mathcal{I}_3$
may produce desirable coverage probabilities at the 90\% level,
they usually have under-coverage probabilities at the 95\% and 99\%  levels.
For example, their under-coverages can be  as large as around 3\%
in Scenario A and $\nu_0=200$.
Although $\mathcal{I}_2$ and $\mathcal{I}_3$  have
 acceptable two-sided coverage probabilities in Scenarios B and D,
their  lower limits  often produce under-coverage probabilities,
while their upper limits often produce over-coverage probabilities.
These phenomenons are more obvious in Scenarios A and C
with $\nu_0 = 200$.
When the population size increases from 200 to 400,
the  one-sided and two-sided coverage probabilities of all the three interval estimators
become more and more accurate.
This is probably because more individuals are observed 
and thus the distributions of the empirical likelihood ratio statistic and pivotal statistics
are closer to their limiting distributions.

\begin{figure}
\begin{center}
\makebox{
\includegraphics[width=0.25\textwidth]{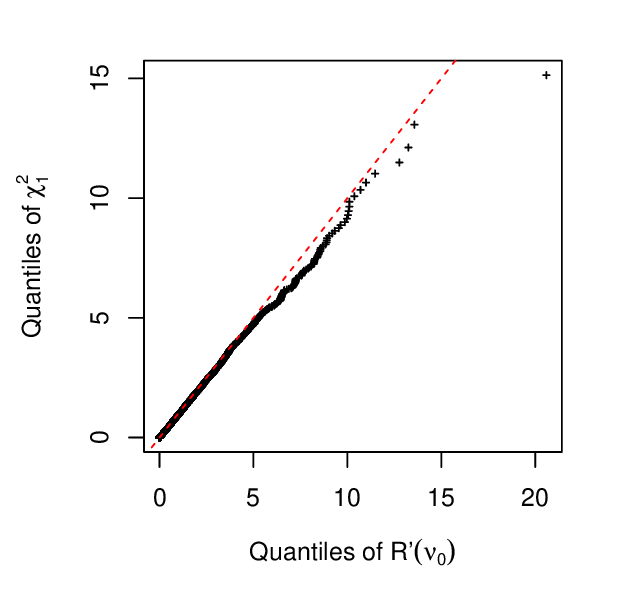}
\includegraphics[width=0.25\textwidth]{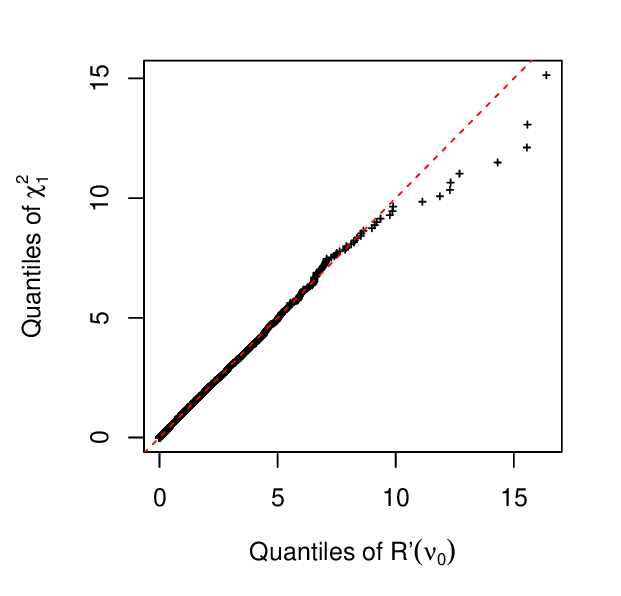}
\includegraphics[width=0.25\textwidth]{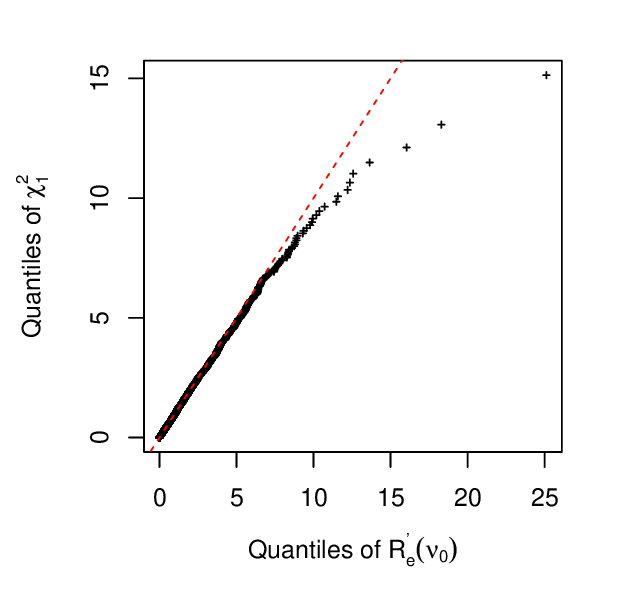}
\includegraphics[width=0.25\textwidth]{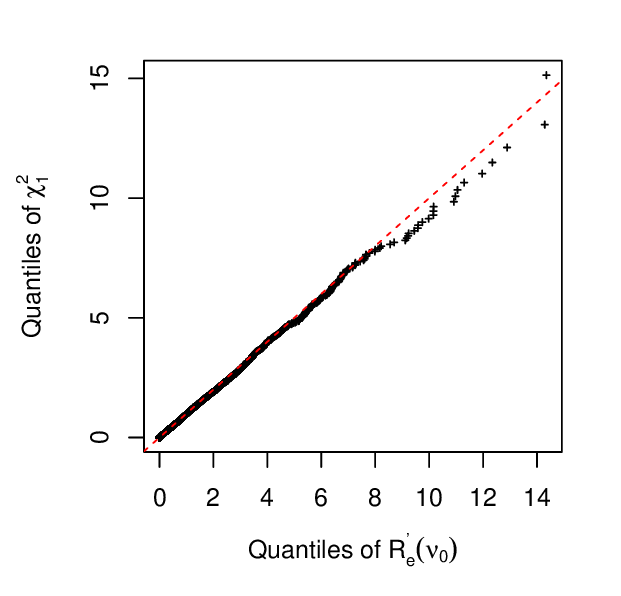}

}
\\
\makebox{
\includegraphics[width=0.25\textwidth]{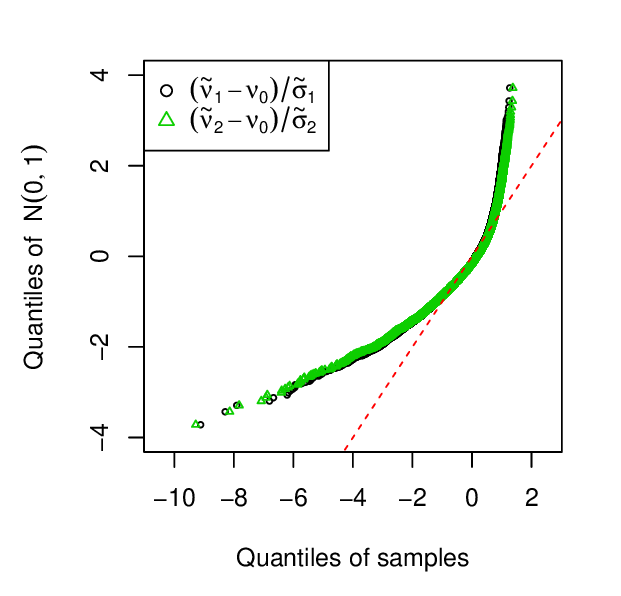}
\includegraphics[width=0.25\textwidth]{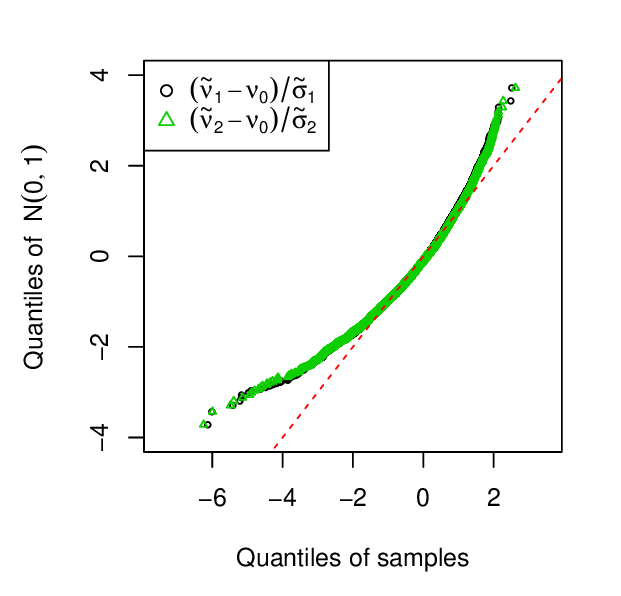}
\includegraphics[width=0.25\textwidth]{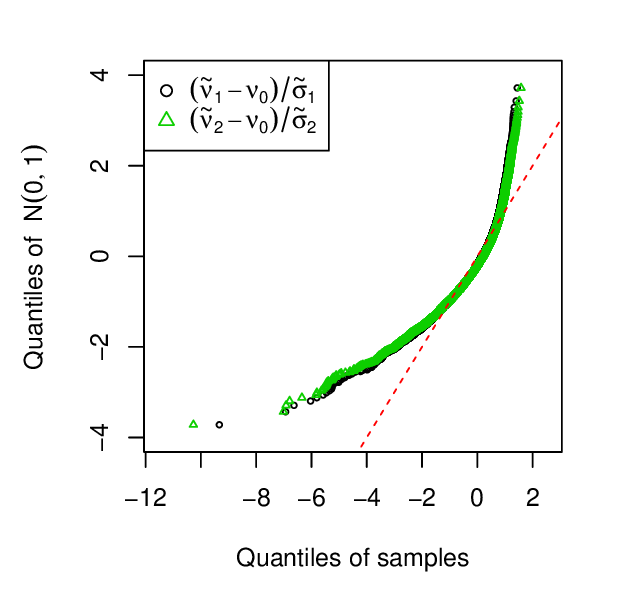}
\includegraphics[width=0.25\textwidth]{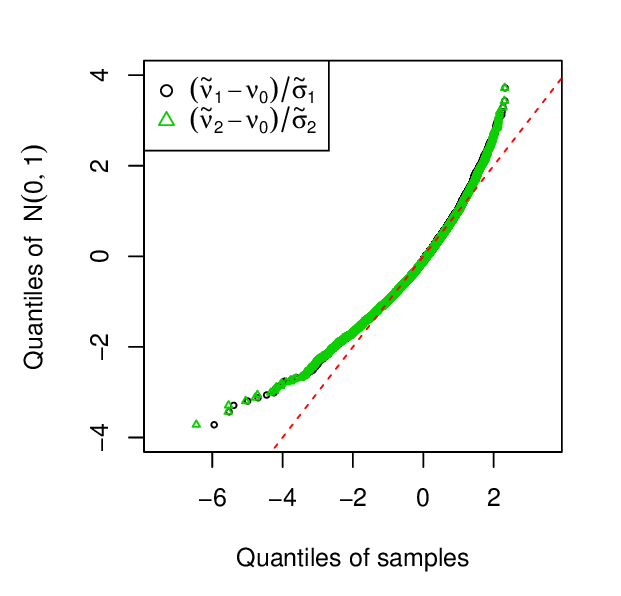}

}
\end{center}
 \caption{ \label{figure-QQ-200}	QQ-plots of $R'(\nu_0)$ and $R_e'(\nu_0)$ (row 1),  
 and two pivotal statistics (row 2)
 in Scenarios  A-D with $\nu_0=200$ (columns 1-4).}
 \end{figure}

{\it Comparison of QQ-plots} \quad
We display in
Figure \ref{figure-QQ-200}
the QQ-plots of
the proposed likelihood ratio statistic $R'(\nu_0)$ or $R'_e(\nu_0)$ versus its limiting $\chi^2_1$ distribution,
and  the two pivotal statistics $(\tilde\nu_1 - \nu_0)/\tilde\sigma_1$
and $(\tilde\nu_2 - \nu_0)/\tilde\sigma_2$
versus their limiting distribution $N(0,1)$ when $\nu_0=200$.
We find that the findings in our interval comparison  can be well
explained by these QQ-plots.

We can clearly see that the finite-sample distributions of $R'(\nu_0)$
are always much closer to its limiting $\chi^2_1$ distribution
than those of $(\tilde\nu_1 - \nu_0)/\tilde\sigma_1$ and $(\tilde\nu_2 - \nu_0)/\tilde\sigma_2$
to their limiting distribution $N(0,1)$.
This explains why the likelihood ratio confidence interval $\mathcal{I}$ has more accurate coverage probabilities
than the two Wald-type  confidence intervals $\mathcal{I}_2$ and $\mathcal{I}_3$.
The coverage probabilities of $\mathcal{I}_2$ and $\mathcal{I}_3$ are very close to each other in all scenarios
due to the almost coincidence of the distributions of their corresponding pivotal statistics.
In addition, the quantiles of these two pivotal statistics are generally
smaller than the standard normal quantiles.
This explains why the  lower limits of the two Wald-type confidence intervals
have sever under-coverages while their upper limits have severe over-coverages.

In summary, if we discard the missing data directly,
the complete-case estimator $\tilde\nu_3$ is unacceptably biased.
However, the proposed  method in this paper
perfectly corrects its bias and produces desirable point and interval estimators
for the abundance.
The proposed estimator $\hat\nu$ usually has less bias and RMSE
than \cite{lee2016estimation}'s two estimators $\tilde\nu_1$ and $\tilde\nu_2$.
The proposed interval estimator
always exhibits better performance than
 the two Wald-type interval estimators $\mathcal{I}_1$ and $\mathcal{I}_2$
 in terms of one-sided and two-sided coverage accuracies.

\section{ Real data }
We illustrate the proposed estimation method by analyzing
the bird species yellow-bellied prinia collected in Hong Kong,
which was analyzed by \cite{lee2016estimation}.
There are 163 distinct birds ever captured on 17 occasions
during an capture-recapture experiment.
We consider two covariates,  fat index (denoted by $X^*$)
and  tail length (denoted by $Y^*$).
The variable $X^*$  is completely observed,
and is defined to be 1 if a bird is classified as fat and 0 otherwise.
The variable  $Y^*$ takes continuous values and is subject to missing with
missing rate   around 25\%.
We assume that the capture probability  satisfies  \eqref{logistic-model}
 with $Z^*=(1,X^*,Y^*)^\T$ (Model 1)
 or $Z^*=(1, Y^*)^\T$ (Model 2).

Table \ref{table-realdata} tabulates the point estimates of $\beta$ and $\nu$,
and 95\% confidence intervals of $\nu$ based on
the proposed  method (Proposed),
the inverse probability weighting  method (IPW),
the multiple imputation  method (MI) and
the complete-case-based method (CC).

We  see that  under both Models 1 and 2,
the first three methods produces
very close estimates of $\beta$ and $\nu$,
which coincides with the observations in our simulation study.
Meanwhile the  CC estimate of $\nu$ is clearly downward biased and unreliable
according to our simulation study.
For interval estimation under the two models,
the lower limits of the two Wald-type interval estimates based on IPW and MI
are no greater than 133, which is even less than 163,
the number of observations in the data.
This is clearly unreasonable.
By contrast,   the proposed interval estimates are more reasonable,
and is expected to have desirable coverage probability,
according to our simulation study.

To choose a better model from Models 1 and 2,
we test $H_0:\beta_{X^*}=0$ under Model 1, where $\beta_{X^*}$
is the coefficient of $X^*$ in the capture probability model. 
Note that $\beta_{X^*}\neq 0$ implies that Model 1 is more appropriate. 
The empirical likelihood ratio test statistic for testing $H_0:\beta_{X^*}=0$ under Model 1
is 8.13. Calibrated by the $\chi^2_1$ limiting distribution, the p-value is around 0.4\%, 
which implies that we have very strong evidence to reject $H_0:\beta_{X^*}=0$. 
Hence Model 1 is more appropriate.
Under this model,
we conclude that there are around 771 yellow-bellied prinia birds in total,
with a 95\% confidence interval [449, 1981].

\begin{table*}[h]
\centering
\caption{\label{table-realdata}
Point estimates of $\nu$ and $\beta$,
and 95\% confidence intervals of $\nu$.
}

\begin{tabular}{ccccc}
\toprule
Model &Method & Estimate of $\nu$& Confidence interval of $\nu$&
    Estimate of $\beta$ \\
\midrule    
\multirow{4}{*}{Model 1} &
Proposed 		& 770  	&  $\mathcal{I}=[449, 1980]$    	&  $(-10.6670, 1.0141, 0.0832)$ \\
&IPW		& 784 	&  $\mathcal{I}_1=[80, 1488]$ 	&  $(-10.7353, 1.0222, 0.0839)$ \\
&MI	    		& 778 	&  $\mathcal{I}_2=[95, 1461]$ 	&  $(-10.6772, 1.0250, 0.0832)$  \\
&CC 		& 406  	&  $\mathcal{I}_3=[258, 916]$ 	&  $(-10.0429, 0.7096, 0.0805)$  \\
\midrule
\multirow{4}{*}{Model 2}&
Proposed  	& 608  	&  $\mathcal{I}=[395, 1311]$ 	&  $(-10.2149, 0.0865)$ \\
&IPW		& 617 	&  $\mathcal{I}_1=[133, 1100]$ &  $(-10.3494, 0.0881)$ \\
&MI 			& 617 	&  $\mathcal{I}_2=[124, 1110]$ 	&  $(-10.3769, 0.0884)$  \\
&CC 		& 362 	&  $\mathcal{I}_3=[244, 729]$ 	&  $( - 9.6187, 0.0816)$  \\

\bottomrule
\end{tabular}
\end{table*}

\section{Discussion}
In this paper, we have proposed a full likelihood method for the abundance estimation
in discrete time capture-recapture model when the covariates are missing at random.
We  have established its large sample property. 
Simulation studies also showed that 
the proposed maximum empirical likelihood estimator has smaller mean square error and 
the empirical likelihood ratio confidence interval is more 
reliable
than those
based on the inverse probability weighting and multiple imputation methods.

To cope with variations in the capture probability,
\cite{otis1978statistical} introduced eight models,
 $M_0, M_t, M_b$, $M_h, M_{tb}, M_{th}, M_{bh}$ and $M_{tbh}$,
according to whether
the capture probability is influenced by
capture occasion or time (t),
behavioral response (b) and heterogeneity between individuals (h).
This paper concentrates on the $M_h$ model.
The proposed method can  be extended to  models such as
$M_{th}, M_{bh}$ and $M_{tbh}$. We leave it as future research. 

\section*{Acknowledgement}
Dr.~Liu was  supported by the National Natural Science Foundation   of China
(Numbers 11771144, 11501354, and  11501208), the
Chinese Ministry of Education 111 Project under Grant number B14019,
and  the Program of Shanghai Subject Chief Scientist under Grant number 14XD1401600.
Dr.~Li was supported in part by  the Natural Sciences and
Engineering Research Council of Canada grant number  RGPIN-2015-06592.

\bibliography{mybib}
\bibliographystyle{chicago}

\end{document}